\documentclass[twocolumn,10pt,a4paper,sort&compress,aps,pre,showpacs,superscriptaddress]{revtex4-2}

\usepackage[english]{babel}

\usepackage[caption=false]{subfig}
\addto\captionsenglish{}

\usepackage{overpic}
\usepackage{amsmath,amssymb,amsthm}
\usepackage{commath}
\usepackage{graphicx}% Include figure files
\usepackage{dcolumn}% Align table columns on decimal point
\usepackage{bm}% bold math
\usepackage{hyperref}% add hypertext capabilities
\usepackage{etoolbox} % for \appto
\usepackage{siunitx}
\usepackage{multirow}
\usepackage[capitalise,nameinlink]{cleveref}
\newcommand{\aref}[1]{\hyperref[#1]{Appendix}}
\usepackage{blkarray}

\usepackage[normalem]{ulem}
\usepackage{comment}
\usepackage[]{color}
\newcommand \bl{\color{black}}

\renewcommand{\vec}[1]{\bm{#1}}

\begin{document}

\title{Self-organized swimming with odd elasticity}

\author{Kenta Ishimoto}
\email{ishimoto@kurims.kyoto-u.ac.jp}
\affiliation{Research Institute for Mathematical Sciences, Kyoto University, Kyoto 606-8502, Japan}

\author{Cl\'{e}ment Moreau}
\email{cmoreau@kurims.kyoto-u.ac.jp}
\affiliation{Research Institute for Mathematical Sciences, Kyoto University, Kyoto 606-8502, Japan}

\author{Kento Yasuda}
\email{yasudak@kurims.kyoto-u.ac.jp}
\affiliation{Research Institute for Mathematical Sciences, Kyoto University, Kyoto 606-8502, Japan}

\date{\today}

\begin{abstract}
We theoretically investigate self-oscillating waves of an active material, which have recently been introduced as a non-symmetric part of the elastic moduli, termed odd elasticity. Using Purcell's three-link swimmer model, we reveal that an odd-elastic filament at  low Reynolds number can swim in a self-organized manner and that the time-periodic dynamics are characterized by a stable limit cycle generated by elastohydrodynamic interactions. Also, %motivated by shape fluctuation originating from internal molecular motors,
we consider a noisy shape gait and derive a swimming formula for a general elastic material in the Stokes regime with its elasticity modulus being represented by a non-symmetric matrix, demonstrating that the odd elasticity produces biased net locomotion from random noise.
 %under a noisy shape gait and theoretically demonstrate the odd elasticity produces biased net locomotion from random noise.
%Abstract must be no more than 600 characters
\end{abstract}

\maketitle

\section{Introduction}
%{\it Introduction.--}
Swimming is a physical outcome of fluid-structure interactions driven by the internal activity of a material. %In particular, time-periodic wave-like beating is ubiquitous in nature, as seen from elastic filaments of cilia and flagella of microorganisms and spermatozoa \cite{Wan2018, Man2020, Lauga2020, Velho2021, Gaffney2021} to undulatory motions of aquatic animals, including the oscillatory fins of dolphins and dinosaurs \cite{Gazzola2015, Smits2019, Ibrahim2020, Heydari2021}. 
% Also wave-like beating has been employed in artificial swimmers from microscopic filaments to fish-like swimming robots \cite{duroure2019}
 In particular, time-periodic wave-like beating is ubiquitous both in {\bl biological and artificial swimmers}, as seen from elastic filaments of microorganisms, spermatozoa, {\bl and micro-actuators} %Wan2018, 
 \cite{DuRoure2019, Man2020, Lauga2020, Velho2021, Gaffney2021} to undulatory motions of aquatic animals {\bl and fish-like robots with oscillatory fins} \cite{Gazzola2015, Smits2019, Duraisamy2019, Heydari2021}. %Ibrahim2020,
Recently, \citet{Scheibner2020} proposed a term, {\it odd elasticity}, which refers to anti-symmetric components of material elastic moduli. This breaks the Maxwell-Betti reciprocity and can cause self-oscillation. The odd elasticity may emerge from non-energy-conserving microscopic interactions in an active material and has gathered intensive attention in the field of non-equilibrium and active matter physics \cite{Scheibner2020b, Bergholtz2021, Brandenbourger2019, Zhou2020, Fruchart2021, Kole2021}.
% The self-induced traveling waves of materials can be formulated by a non-Hermitian operator from quantum to classical systems, and this breaks the Maxwell-Betti reciprocity of the material response \cite{Ashida2020, Shankar2020, Scheibner2020b, Bergholtz2021, Brandenbourger2019, Zhou2020, Fruchart2021, Kole2021}. For an elastic material, such self-oscillation has recently been captured by anti-symmetric components of material elastic moduli, termed as odd elasticity, which emerges from non-energy-conserving microscopic interactions in an active medium \cite{Scheibner2020}. In an overdamped system with linear elasticity, the non-symmetric elasticity tensor yields non-real eigenvalues that correspond to the self-induced wave, for which non-reciprocal elastic deformation will be generated.
%Material activity is, in general, generated by internal chemical reactions such as molecular machines (e.g., kinesin, dynein, and myosin) and particular enzymes \cite{Ariga2018, Brown2019, Mugnai2020, Yasuda2021b, Ghosh2021}, whose shapes typically fluctuate over time. 

To achieve swimming at the microscale, it is well known as the {\it scallop theorem} that one needs to deform in a non-reciprocal manner in the fluid \cite{Purcell1977}. This non-reciprocal deformation is theoretically formulated by the gauge field theory and represented as a closed loop in shape space with non-zero area \cite{Shapere1989}.

%For swimming at the microscale, in particular, these non-reciprocal deformations are required for generating net locomotion to overcome the time-reversal property of the Stokes flow, or the scallop theorem \cite{Purcell1977}, as theoretically formulated by the gauge field theory \cite{Shapere1989}. }
Moreover, microscopic propulsion is often accompanied by fluctuations from the internal motors or environmental stochasticity.
Recently, \citet{Yasuda2021} analyzed a three-sphere swimmer linked by two odd-elastic arms under thermal fluctuations and showed that the swimmer can exhibit directed locomotion from random noise as a statistical average. This result implies that non-reciprocal deformation for microswimming may be induced by the odd elasticity that can cause non-reciprocity of elastic response. %achieved by the non-reciprocity of elastic response caused by the odd elasticity.}
%suggesting that an odd elastic material can propel itself by its self-induced non-reciprocal deformation. 
{\bl Note that these two reciprocities are conceptually distinct: one is a geometric property of a path, the other classifies the constitutive properties of a mechanical system.}
The sphere model was, however, restricted to one-dimensional motion and was not able to swim in a time-periodic beating fashion.
% In the kinematic model, the loop is given by hand, but in the elastohydrodynamic problem, the trajectory in the shape space is simultaneously determined by the equations of the system.

Therefore, the %first
aim {\bl of this paper} is to extend the  previous swimming theory of odd-elastic material by \citet{Yasuda2021} to planar or higher-dimensional motion as well as to an arbitrary number of dimensions of shape space, and determine whether such a material can self-propel. We also seek universal features for the dynamics of a generalized linear elastic material at low Reynolds number, assuming the odd-elastic modulus as a simple coarse-grained representation of material activity {\bl -- although this representation is not claimed to model or explain biological microswimming by odd elasticity.} 
The shape gait of an active material is, in general, given by %determined as
the solution to an elastohydrodynamics coupling problem, which is non-local and non-linear due to the material geometry even at low Reynolds number. 
Thus,  
as a simple but canonical model, we will first consider a  coarse-grained description of a swimming filament known as Purcell's three-link swimmer \cite{Purcell1977} (Fig.\ref{fig:tls}(a)).  As a minimal model of microswimming with two degrees of freedom, it has been studied to understand various aspects of biological swimmers and artificial robots, such as efficiency, stability, and control \cite{Becker2003, Avron2008, Passov2012, Hatton2013, Giraldi2015, moreau2019local, Zheng2021}. 

Many models of elastohydrodynamic swimming require programmed internal forces that drive the material as an input function \cite{Olson2013, Simons2015, Ishimoto2018a, Gadelha2019}, or further modeling on the internal structure, for example, the flagellar structure and regulation mechanism of molecular motors \cite{Riedel2007, Evans2010, Gadelha2013, Bayly2015}. These models are, therefore, problem-specific in general. In contrast, we will see that the odd-elasticity description leads to autonomous elastohydrodynamics equations. % our second aim in this Letter is to seek universal features for the dynamics of the generalized linear elastic material at low Reynolds number, assuming the odd-elastic modulus as a simple coarse-grained representation of material activity. 
%elastohydrodynamics of an odd-elastic filament as a coarse-grained continuum description of internal activity.Hence,

In the following, we will use the Purcell swimmer model to demonstrate that an odd-elastic filament can swim in a self-organized manner, by which we mean pattern formation in the system far from thermal equilibrium without any programmed driving forces. %exhibit stable swimming in viscous fluid as a self-organized elastohydrodynamic buckling instability.
We will then proceed to consider an odd-elastic material under fluctuations, motivated by biological {\bl and artificial swimmers, including sperm, {\it Chlamydomonas} , and Janus particles} \cite{Ma2014, Wan2014a, Ishimoto2017b, nishiguchi2018}, whose shape gaits are characterized by a noisy limit cycle. % driven by the internal molecular fluctuations. Further, 
 Finally, we will describe a general odd-elastic material with an arbitrary number of degrees of freedom, show that the odd elasticity produces net locomotion from random noise, and derive a swimming formula that provides ensemble-averaged swimming velocity as a coupling of the swimmer gauge field and probability current in the shape space. % We finally demonstrate that the odd elasticity triggers biased net locomotion from a non-biased random noise and geometrical non-linearity can amplify a stochastic noise to merge stable swimming as a limit cycle in the shape space. 

%\section{Model}
\section{Purcell's swimmer with odd elasticity}
\begin{figure}[!t]
\begin{center}
\begin{overpic}[width=8.5cm]{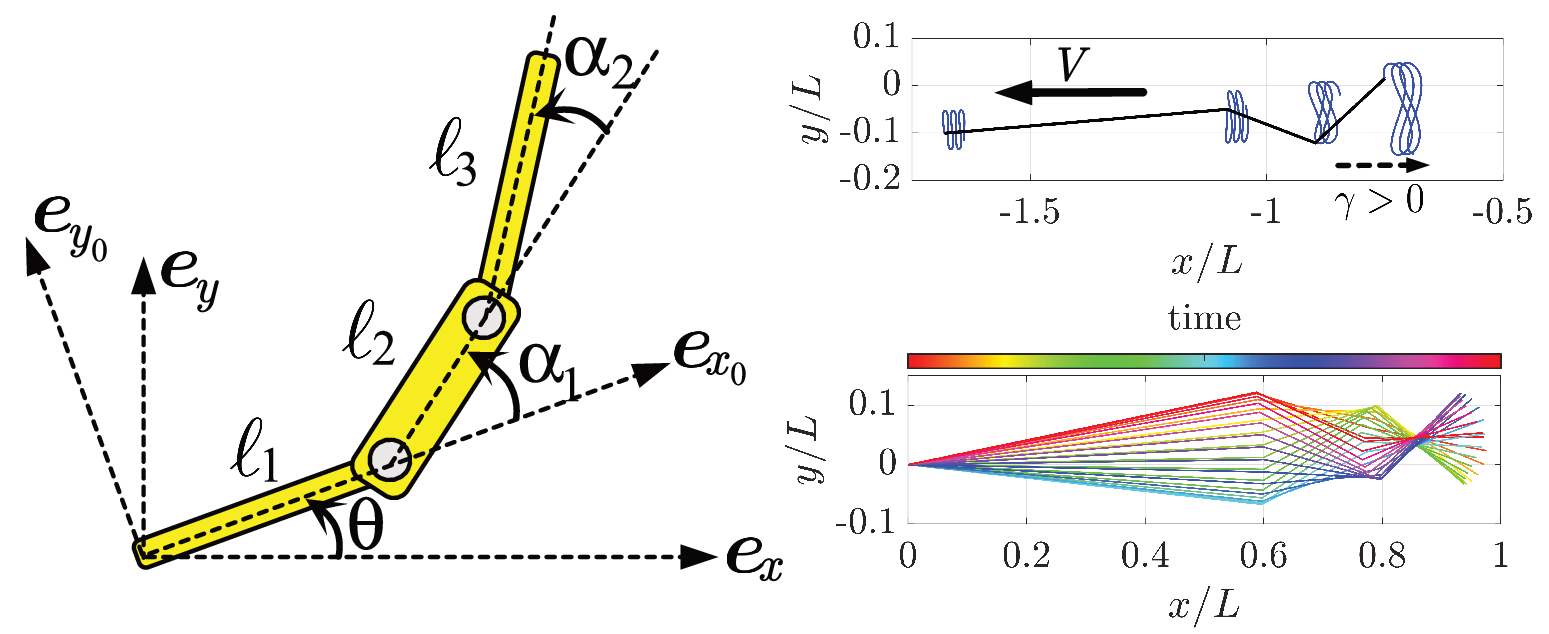}
\put(5,42){(a)}
\put(50,42){(b)}
\put(50,18){(c)}
\end{overpic}
%\begin{overpic}[width=4cm]{fig_tls.eps}\put(2,54){(a)}\end{overpic}~
%\begin{overpic}[width=2.5cm]{fig_tls2.eps}\put(0,90){(b)}\end{overpic}\\
\vspace*{-0.3cm}
\caption{(a) Schematic of Purcell's three-link swimmer. (b) An example trajectory of each end of the three rods in self-organized swimming (in the -$x$ direction) for a pusher swimmer with $\gamma>\gamma_c$. (c) Rod shape of the same swimmer as (b) superimposed with its left-most rod end being translated to the origin. The colors indicate the time increment over one beat period.} \label{fig:tls}
\end{center}
\vspace*{-0.3cm}
\end{figure}

The three-link model swimmer, known as Purcell's swimmer \cite{Purcell1977}, consists of three slender rods of lengths $\ell_1$, $\ell_2$, and $\ell_3$ connected by two hinges, as shown in Fig. \ref{fig:tls}(a), which also introduces lengthscale $L=\ell_1+\ell_2+\ell_3$. We denote the position of the end of the first rod as $(x, y)$ and the angle from the $x$ axis as $\theta$. The relative angles at the two hinges are denoted as $\alpha_1$ and $\alpha_2$. We assume the hinges are elastic \cite{Moreau2019}, and linearly related to the relative angles so that the $\bm{e}_z$ component of the elastic torque is given by $T_\alpha=K_{\alpha\beta}\alpha_\beta$, with Greek indices denoting the degrees of freedom for the shape, as $\alpha, \beta=\{1, 2\}$. This linear elastic hinge at the linkage may be regarded as a coarse-grained representation of the Euler-Bernoulli constitutive  relation \cite{Moreau2018, Walker2019d, walker2020efficient}.  To ensure that the object relaxes to an equilibrium configuration in the absence of odd elasticity, we assume the matrix $K_{\alpha\beta}$ to be positive-definite. Moreover, following previous studies \citep{Scheibner2020, Yasuda2021}, we consider a simple form of the elasticity matrix as 
\begin{equation}
    K_{\alpha\beta}=\kappa_e\delta_{\alpha\beta}+\kappa_o\epsilon_{\alpha\beta}
\label{eq:M01},
\end{equation}
where $\kappa_e$ and $\kappa_o$ are the even and odd-elastic moduli, $\delta_{\alpha\beta}$ is the Kronecker delta, and $\epsilon_{\alpha\beta}$ is the two-dimensional anti-symmetric tensor. We will henceforth write the ratio of the two elastic moduli as $\gamma= \kappa_o /\kappa_e$.  Note that the $\kappa_e$ is assumed to be positive but $\kappa_o$ may be an arbitrary real number.

To show the equations of swimming dynamics, which obey the steady Stokes equations of low-Reynolds-number flow, we introduce the body-fixed coordinates $\{\bm{e}_{x0}, \bm{e}_{y0}, \bm{e}_{z0}\}$, whose origin is located at the end of the first rod. 
%{\it Elastohydrodynamic equations.--}
Using the resistive force theory and force- and torque-free condition for the swimmer, its dynamics are given in the body-fixed coordinates by
\begin{equation}
    -{\bf M}(\alpha_1, \alpha_2)\dot{\bm{z}}={\bf L}\bm{z}
\label{eq:M02},
\end{equation}
where $\bm{z}=(x_0, y_0, \theta, \alpha_1, \alpha_2)^\textrm{T}$ and the dot represents the time derivative. The $5\times 5$ matrix ${\bf M}$ can be taken as being symmetric, positive-definite, and dependent only on the shape parameters, $\alpha_1$ and $\alpha_2$, {\bl with further description of its properties being provided in  Appendix \ref{App:MatM}.} We hereafter use Roman indices for the rigid motion in the physical space such as $i,j=\{1,2,3\}$ to distinguish them from the Greek indices for the shape space. The matrix ${\bf L}$ includes the elasticity matrix such that $L_{3+\alpha,3+\beta}=K_{\alpha\beta}$ and the other components of ${\bf L}$ are zero.
 From the matrix structure of the dynamics (\ref{eq:M02}), the solution is formally obtained by inverting the matrix ${\bf M}$. Letting ${\bf N}={\bf M}^{-1}$, we can decompose the equations into those for rigid motion and shape deformation, with $\bm{z}_0=(x_0, y_0, \theta)^\textrm{T}$ and $\bm{\alpha}=(\alpha_1, \alpha_2)^\textrm{T}$, as
\begin{equation}
\dot{\bm{z}}_0=-{\bf P}{\bf K}\bm{\alpha}
\textrm{~and~}
\dot{\bm{\alpha}}=-{\bf Q}{\bf K}\bm{\alpha}
    \label{eq:M11b},
\end{equation}
where $P_{i\alpha}=N_{i, 3+\alpha}$ and $Q_{\alpha\beta}=N_{3+\alpha, 3+\beta}$.
The second equation of (\ref{eq:M11b}) is closed with respect to the shape angles, whereas the first equation has an alternative form,
%\begin{equation}
$\dot{\bm{z}}_0= {\bf P}{\bf Q}^{-1}\dot{\bm{\alpha}}
    \label{eq:M12}
$%\end{equation}
,that is not explicitly dependent on the elastic matrix and identical to the kinematic problem. %With expanding this form around the equilibrium up to the second order, we recover (\ref{eq:M16}). Of note, the symbolic elastohydrodynamic equations (\ref{eq:M02}) and (\ref{eq:M11b}) can be extended to a general odd-elastic material with arbitrary degrees of freedom.

%\section{Swimming dynamics}
%\subsection{Existence of limit cycles for the three-link swimmer}

\section{Self-organized swimming as a stable limit cycle}
\begin{figure}[!t]
%\begin{center}
%\begin{overpic}[width=4.2cm]{fig_3_1_1_500_shape1.eps}\put(-4,32){(a)}\end{overpic}~
%\begin{overpic}[width=4.2cm]{fig_3_1_1_500_shape2.eps}\put(-4,32){(b)}\end{overpic}\\
%\vspace{1em}
\vspace*{+0.3cm}
\hspace*{-0.3cm}
%\begin{overpic}[width=8.8cm]{bifurcation_both_1.eps}\put(-2,43){(c)}\put(49,43){(d)}\end{overpic}
\begin{overpic}[width=9.0cm]{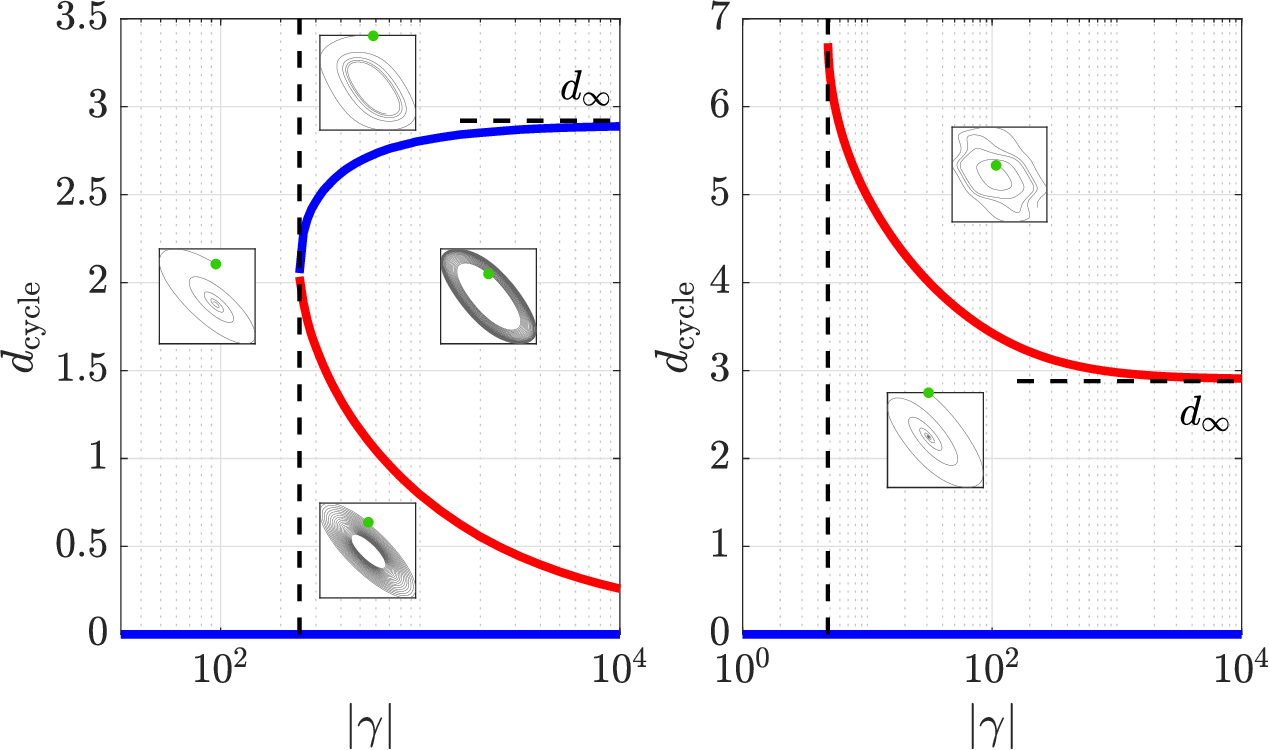}
\put(-1,57){(a)}
\put(50.5,57){(b)}
\put(22,59){Pusher}
\put(73.5,59){Puller}
\put(22,5.6){$\gamma_c$}
\put(64,5.6){$\gamma_c$}
\end{overpic}
%\begin{overpic}[width=4.2cm]{bifurcation.eps}\put(-4,68){(d)}\end{overpic}
\vspace*{-0.5cm}
\caption{Bifurcation diagrams for (a) a pusher swimmer and (b) a puller swimmer with sample trajectories in the shape space $(\alpha_1,\alpha_2)$ shown in insets. The diameter of the cycle orbit, $d_{\mathrm{cycle}}$, is plotted as a function of $|\gamma|$. The equilibrium configuration is always linearly stable for a finite $|\gamma|$ for both swimmers. The pusher swimmer dynamics exhibits stable (blue) and unstable (red) limit cycles  above a critical value of $\gamma$, at which a semi-stable limit cycle bifurcation occurs. For the puller swimmer, in contrast, the limit cycle is always unstable.  The diameters of the stable cycle in case (a) and of the unstable cycle in case (b) both converge to the same value $d_{\infty}$ as $| \gamma | \rightarrow \infty$ due to  time-reversal symmetry. Initial configurations are shown by a green dot in each inset. }
\label{fig:bifurcation}
%\end{center}
\end{figure}
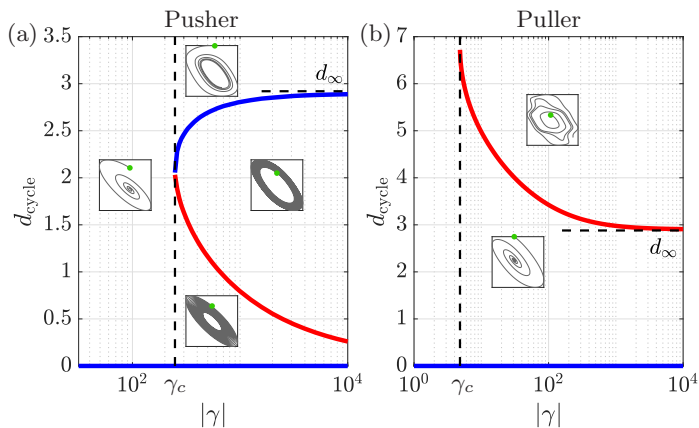

Numerical explorations revealed that the Purcell swimmer can swim in a self-organized manner%, by which we mean pattern formation in the system far from thermal equilibrium without any explicit driving forces.
Such a periodic locomotion only occurs  when the swimmer shape has fore-aft asymmetry, i.e., $l_1\neq l_3$. In Fig. \ref{fig:tls}(b), sample trajectories of the ends of the rods are shown with stable periodic shape gait [Fig. \ref{fig:tls}(c)]. With the rod lengths $\ell_1>\ell_3\sim\ell_2$ and $\gamma>0$, the object can swim towards the left end (negative $\bm{e}_x$ axis) as the beating wave travels down towards the right [Fig. \ref{fig:tls}(b)]. The right-most rod vigorously oscillates like a {\it pusher} swimmer, such as sperm cells. With the reversed sign of odd elasticity ($\gamma<0$), the swimming direction is also reversed with its oscillatory part being ahead of the longest rod, like a {\it puller} swimmer, such as {\it Leishmania} \cite{Walker2019}. Of note, the pusher or puller behavior of the swimmer, as well as the swimming direction, depends not only on the sign of $\gamma$, but also on the swimmer's geometry. {\bl In the puller case, we did not observe stable swimming}, [see Fig. \ref{fig:bifurcation}(b)], in the sense that the swimmer either exhibits unstable trajectories with the $\alpha$ angles amplifying until the links overlap, or eventually reaches the zero equilibrium. {\bl Here,} the decay to this equilibrium becomes notably slow, scaling roughly with $\mathcal{O}(1/\sqrt{t})$ as $| \gamma |$ tends to infinity.

%\begin{comment}
%\end{comment}

\begin{comment}
\begin{figure*}[!t]
\begin{center}
\begin{overpic}[width=18cm]{fig2old.eps}\put(0,34){(a)}\put(39,34){(b)}\put(76,34){(c)}\put(76,16){(d)}\end{overpic}
\caption{Bifurcation diagrams for (a) a pusher and (b) a puller swimmer. The diameter of the cycle orbit, $d_{\text{cycle}}$, is plotted as a function of $\gamma$. The equilibrium configuration is always linear stable for a finite $\gamma$ for both swimmers. The pusher swimmer dynamics exhibit stable (blue) and unstable (red) limit cycles above a critical value of $\gamma$, at which semi-stable limit cycle bifurcation occurs. The puller swimmer, in contrast, the limit cycle is always unstable.
(c) A trajectory in the shape space of a self-organized swimmer under noise. A pusher swimmer with $\gamma<0$ was initially at rest with a straight configuration, but once the shape exceeds the inner unstable limit cycle (red broken line) it converges to the outer stable limit (blue dashed line), yielding self-organized periodic swimming. See also Supplemental Movie. (d) An illustration of the steady probability distribution shown by color contour and the probabilistic current vector in the shape space depicted by blue arrows.}
\label{fig:bifurcation}
\end{center}
\vspace*{-0.75cm}
\end{figure*}
\end{comment}

%\subsection{Stability analysis}
%{\it Stability of equilibrium.--}
We now proceed to a bifurcation analysis of the elastohydrodynamic dynamical system. Around the equilibrium straight configuration, the dynamics in the shape space (\ref{eq:M11b}) is linearized, with  ${\bf Q}(\bm{\alpha}=\bm{0})$ denoted by ${\bf Q}_0$ and ${\bf \Gamma}={\bf Q}_0 {\bf K}$, to
\begin{equation}
\dot{\bm{\alpha}}=-{\bf \Gamma}\bm{\alpha},    
\label{eq:S21}.
\end{equation}
Noting that the matrix ${\bf Q}_0$ is symmetric, we obtain %found to be symmetry by direct calculation. 
the eigenvalues of ${\bf \Gamma}$ as
\begin{equation}
\lambda=\frac{\kappa_e }{2}\left[ \textrm{Tr}{\bf Q}_0 \pm 
\sqrt{(\textrm{Tr}{\bf Q}_0)^2-4 (1+\gamma^2)\,\textrm{det}{\bf Q}_0 }\right]    
\label{eq:S22}.
\end{equation}
By virtue of the positive-definiteness of the matrix, ${\bf Q}_0$, the real part of the eigenvalues are found to be all positive, which therefore implies that the dynamics around the straight equilibrium configuration is always linearly stable [Fig. \ref{fig:bifurcation}(a,b)]. %The equilibrium fixed point, however, becomes center if we consider the zero $\kappa_e$ limit or $\gamma\rightarrow\infty$.

%\subsection{Semi-stable limit cycle bifurcation and pusher-puller duality}
%{\it Semi-stable limit cycle bifurcation.--}
We further analyzed the bifurcation structure and found that the system exhibits semi-stable limit cycle bifurcation at a certain $\gamma=\gamma_c$ when the swimmer self-propels as a pusher [Fig. \ref{fig:bifurcation}(a)]. In the phase space, the outer stable limit cycle contains an unstable limit cycle inside, while the origin of the phase space is a stable fixed point [Fig. \ref{fig:noise}(a)]. 
Since a pusher-type stroke generates {\bl extensional} flow along the swimming direction, the rod receives contractile force from the fluid as its reaction. When $\gamma$ exceeds the critical value $\gamma_c$, this contractile force can {\bl balance the elastic relaxation.} %of the even-elastic force that relaxes the rod to a straight configuration.
This morphological transition generated by the contractile forces is similar to flagellar buckling dynamics \cite{Kumar2019, Walker2019d}, but here the bifurcation occurs in a self-organized manner. We note, as shown later in this {\bl paper} [Fig. 3(a)], that the stable limit cycle is reachable from straight configuration under a finite amount of noise, even though the straight configuration is linearly stable. For a puller swimmer, in contrast,  the self-induced oscillation acts as an extensile force on the rod as a reaction to the fluid. %propulsion force becomes extensile
The increase of $\gamma$, therefore, accelerates the elastic relaxation and the stable limit cycle for periodic swimming cannot be realized [Fig. \ref{fig:bifurcation}(b)]. These bifurcation structures are robustly observed in a large range of parameters. {\bl Further discussions are provided in Appendix \ref{App:geom}.} When $|\gamma|\rightarrow\infty$, the time-reversal symmetry of the Stokes equations implies that the dynamics in the shape space are invariant under the change of variables $(t, \ell_1, \ell_3, \alpha_1, \alpha_2)\mapsto (-t, \ell_3, \ell_1, -\alpha_2, -\alpha_1)$. Thus, at this limit, the stability of a  pusher-type rod is opposite from that of a corresponding puller-type rod \cite{Ishimoto2013}. Hence, the swimmer with fore-aft symmetric geometry, i.e., $l_1=l_3$, follows a closed trajectory in the shape space at this limit, while the dynamics only possess a stable fixed point at the straight configuration at a finite $\gamma$. 

As discussed above, the stable limit cycle is only enabled by the broken fore-aft symmetry of the system. Of particular note, this can also be achieved by an asymmetric boundary condition such as one end of the rod being fixed [we enforce $(x,y)=(0,0)$], instead of a geometric asymmetry ($\ell_1 \neq \ell_3$). %This accounts for ciliary stroke rather than flagellar swimming. 
Because of this fore-aft symmetry break induced by the fixed left end, the semi-stable limit cycle bifurcation occurs at a certain level of $\gamma>0$, as in the free-swimming pusher case, even if $\ell_1 = \ell_3$, and both in clamped ($\theta$ fixed to 0) and hinged (unconstrained $\theta$) boundary conditions. In the puller case ($\ell_1 = \ell_3$, $\gamma < 0$), no stable limit cycle is observed -- just like the free swimmer, the deformation pattern then either relaxes to straight equilibrium or exhibit unstable behavior.

%\section{Dynamics with noise}
\section{Noisy swimming around an equilibrium shape}
Motivated by observations of biological {\bl and artificial} swimmers \cite{Ma2014, Wan2014a, Ishimoto2017b, nishiguchi2018}, we consider swimming under an active Gaussian fluctuation inside the swimmer, and the dynamics around an equilibrium shape are then given by
\begin{equation}
\dot{\bm{\alpha}}=-{\bf \Gamma} \bm{\alpha} +\bm{\xi}(t)
\label{eq:F01},
\end{equation}
where ${\bf \Gamma}$ is positive-definite around the equilibrium; the Gaussian noise satisfies $\langle \xi_\alpha\rangle =0$; and
$\langle \xi_{\alpha}(t)\xi_{\beta}(t')\rangle =2D_{\alpha\beta}\delta(t-t')$, where the brackets indicate ensemble average and the diffusion matrix ${\bf D}$ is symmetric and positive-definite. %[(KY) moved the sentences to p.4]

%\begin{comment}
\begin{figure}[!t]
%\begin{center}
\hspace*{-0.7cm}
%\begin{overpic}[width=4.3cm]{plot_noise_113_0.01_3_5b4.eps}\put(-2,68){(a)}\end{overpic}
%\begin{overpic}[width=4.3cm]{fig_1_1_3_300b.eps}\put(-2,68){(b)}\end{overpic}\\
\begin{overpic}[width=10.0cm]{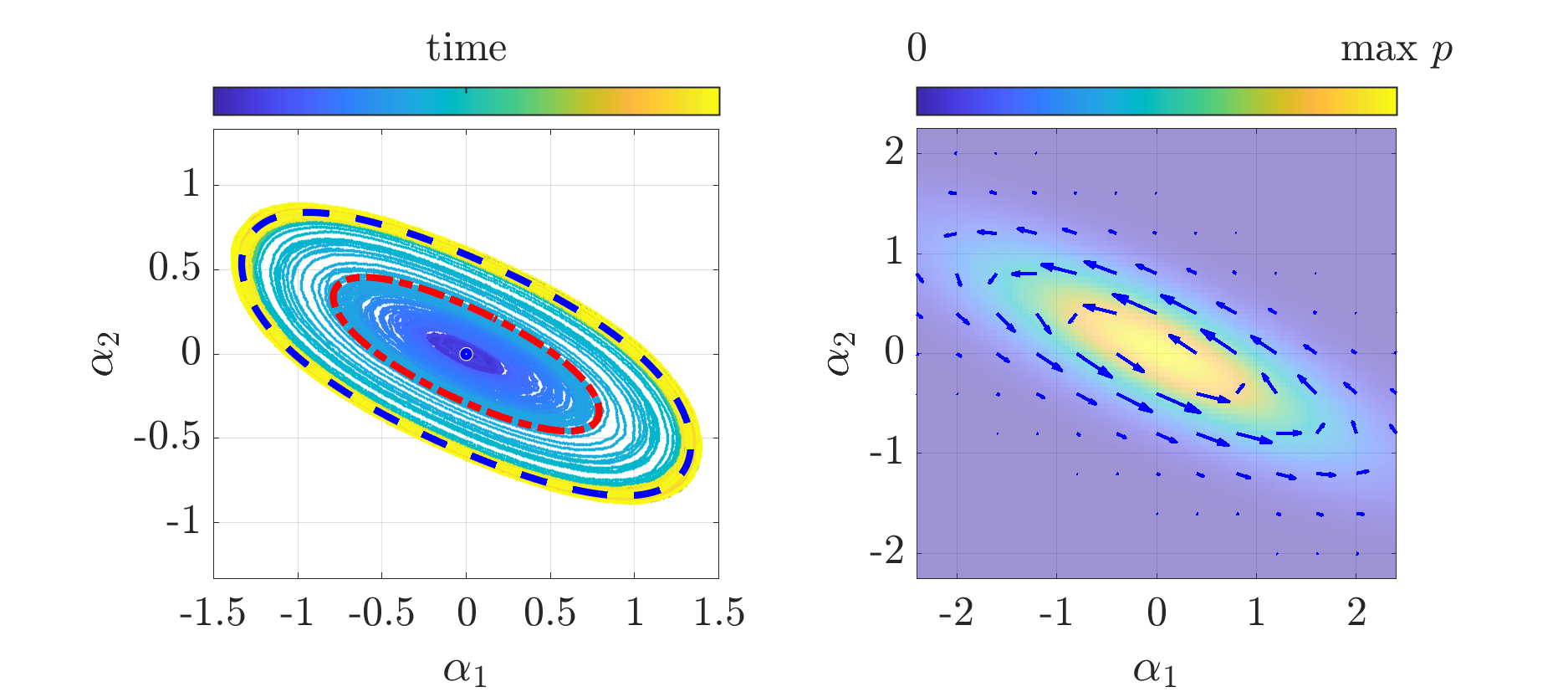}\put(6,40){(a)}\put(50,40){(b)}\end{overpic}
%\begin{overpic}[width=4cm]{plot_noise_113_0.01_3_0.3b_aveloc.eps}\put(-2,68){(c)}\end{overpic}
%\begin{overpic}[width=4cm]{plot_noise_113_0.01_3_0.3b_bveloc.eps}\put(-2,68){(d)}\end{overpic}
\vspace*{-0.4cm}
\caption{(a) Trajectory in the shape space of a self-organized swimmer with noise. The time evolution is shown as a color gradation. A pusher swimmer with $\gamma<0$ was initially at rest with a straight configuration, but once the shape exceeds the inner unstable limit cycle (red broken ellipse), it converges to the outer stable limit (blue dashed ellipse), yielding self-organized periodic swimming. See also Supplemental Movie. (b) Illustration of the steady probability distribution shown by color contour and the probabilistic current vector in the shape space depicted by arrows.  %(c) The averaged areal velocity is plotted in red, whereas the actual evolution is shown in blue. (d) Swimming speed $|\bm{V}|$ is plotted as in (c). 
% These figures will be replaced to fit with other plots. The movie will also be added in supplementary.
}
\label{fig:noise}
%\end{center}
\vspace*{-0.1cm}
\end{figure}
%\end{comment}

Now we rewrite the swimming dynamics using the gauge field formulation \cite{Shapere1989}. The two-dimensional rotation and linear transformation form a two-dimensional Euclidean group \cite{Lapa2014}, and we represent the rigid motion and its generator as
\begin{equation}
    \mathcal{R}=\begin{pmatrix}
    \cos\theta & \sin\theta & x \\
    -\sin\theta & \cos\theta & y \\
    0 & 0 & 1 \\
    \end{pmatrix}\textrm{~and~}
        \mathcal{A}=\begin{pmatrix}
    0 & \dot{\theta} & \dot{x}_0 \\
    -\dot{\theta} & 0 & \dot{y}_0 \\
    0 & 0 & 0 \\
    \end{pmatrix}
    \label{eq:M13},
\end{equation}
which satisfy $\dot{\mathcal{R}}=\mathcal{R}\mathcal{A}$.
The shape in the reference frame, $\mathcal{S}(t)$, is obtained from a rotation of that in the body-fixed frame, $\mathcal{S}_0(t)$, because $\mathcal{S}(t)=\mathcal{R}(t)\mathcal{S}_0(t)$.
Further, we rewrite $\mathcal{A}=\mathcal{A}_\alpha\dot{\alpha}_\alpha$, where the matrix $\mathcal{A}_\alpha(\bm{\alpha})$ is only dependent on the shape parameters $\bm{\alpha}$ and corresponds to the gauge potential associated with $\alpha_\alpha$. Time integration of (\ref{eq:M13}) yields the path-ordering expression of the swimming dynamics. With a small deformation, $\alpha_1, \alpha_2\ll 1$, let us expand the gauge potential around $\bm{\alpha}=\bm{0}$ up to the quadratic term. If the deformation is periodic in time, the average swimming velocity is written  \cite{Shapere1989, Avron2008} as
\begin{equation}
  \overline{\mathcal{A}} =\frac{1}{2}\mathcal{F}_{\alpha\beta}\, \overline{\alpha_{\alpha}\dot{\alpha}_{\beta}}
\label{eq:M16},
\end{equation}
where a bar indicates time average and $\mathcal{F}_{\alpha\beta}=\partial_\alpha \mathcal{A}_\beta-\partial_\beta \mathcal{A}_\alpha-[\mathcal{A}_\alpha, \mathcal{A}_\beta]$ is called the curvature tensor of the gauge field or field strength. We use the notation $\partial_\beta:=\partial/\partial \alpha_\beta$ and the brackets denote the commutator, $[\mathcal{A}_\alpha, \mathcal{A}_\beta]:=\mathcal{A}_\alpha \mathcal{A}_\beta- \mathcal{A}_\beta \mathcal{A}_\alpha$.
The gauge field and its strength can also be represented as a third-rank tensor $[\mathcal{A}_\alpha]_{ij}=A_{ij\alpha}$ and a fourth-order tensor $[\mathcal{F}_{\alpha\beta}]_{ij}=F_{ij\alpha\beta}$, respectively.

We then seek to find the average swimming velocity, $\langle \mathcal{A}\rangle$, considering the probability distribution function in the shape space $p(\bm{\alpha},t)$ that obeys the Fokker-Planck equation associated with (\ref{eq:F01}), i.e., $\partial_t \,p +\partial_\alpha j_\alpha=0$ with $\partial_t=\partial/\partial t$ and the probability flux $j_\alpha$, given by $j_\alpha=-\,\Gamma_{\alpha\beta}\,(\alpha_\beta p) -D_{\alpha\beta} (\partial_\beta p)$. When the distribution is steady, the probability flux draws closed loops [Fig. \ref{fig:noise}(d)], for which we can expect that net locomotion will be produced in an averaged manner \cite{Yasuda2021}. Plugging (\ref{eq:F01}) into (\ref{eq:M16}), the average gauge field becomes $\langle \mathcal{A}\rangle=\mathcal{F}_{\alpha\beta}\langle -\Gamma_{\beta\gamma}\alpha_\gamma\alpha_\alpha+ \alpha_\alpha\xi_\beta\rangle$. We then use the relation $\langle \alpha_\alpha\xi_\beta\rangle=D_{\alpha\beta}$ and introduce the shape covariance matrix, $C_{\alpha\beta}=\langle\alpha_\alpha\alpha_\beta \rangle$, which is the solution to the
Lyapunov equation \cite{Weiss2007},
\begin{equation}
{\bf \Gamma C} + {\bf C\Gamma}^\textrm{T}=2{\bf D}    
\label{eq:N06},
\end{equation}
with superscript $\textrm{T}$ denoting transpose of the matrix. Finally, we can derive the average swimming velocity around the equilibrium by introducing a matrix ${\bf J}=\frac{1}{2}({\bf D}-{\bf \Gamma} {\bf C})$ in the form %{\bl ()}
\begin{equation}
\langle \mathcal{A} \rangle =\textrm{Tr}(\mathcal{F}{\bf J})
\label{eq:F03},
\end{equation}
where the trace is taken over the shape components, i.e., $\langle A_{ij}\rangle=F_{ij\alpha\beta}J_{\beta\alpha}$. {\bl Further derivations and physical interpretations are provided in Appendix \ref{App:noise}.}
This formula (\ref{eq:F03}) is a generalization of the deterministic swimming dynamics (\ref{eq:M16}), and the matrix ${\bf J}$ %, which is proven to be anti-symmetric by plugging it into the Lyapunov equation, 
can be physically interpreted as the areal velocity of the probabilistic current in the shape space. 
 By the definition of ${\bf J}$, its transpose becomes $2{\bf J}^{\textrm{T}}={\bf C\Omega}^{\textrm{T}}={\bf C}({\bf C}^{-1}{\bf D}-{\bf \Gamma}^{\textrm{T}})$, and 
from the Lyapunov equation (\ref{eq:N06}), it follows that ${\bf J}^{\textrm{T}}=-{\bf J}$, from which we conclude that ${\bf J}$ is anti-symmetric.

\section{Purcell's swimmer under shape fluctuation}
With straightforward calculations, we can obtain the gauge field strength around the equilibrium, and find that only the components $F_{1312}=-F_{1321}$ are nonzero and the other components are zero. Thus, only swimming along the $x$-axis is possible, if a time-periodic deformation is considered. Let us denote the nonzero components as $F_{13\alpha\beta}=F\epsilon_{\alpha\beta}$, which is given by
\begin{equation}
F=-\ell_1\ell_2\ell_3(\ell_1^2+\ell_3^2+\ell_1\ell_2+\ell_2\ell_3+\ell_3\ell_1)L^{-4}
\label{eq:F11}.
\end{equation}
This form is in agreement with %the one found in
previous studies of Purcell's swimmer \cite{Becker2003, Koens2021}.
Because the shape space is two-dimensional, the shape covariance, which is the solution to the Lyapunov equation \cite{Weiss2007}, may be solved as
\begin{equation}
{\bf C}=\frac{1}{\textrm{Tr}{\bf\Gamma}}\left[ {\bf D}+(\textrm{det}{\bf \Gamma}){\bf \Gamma}^{-1}{\bf D}({ \bf\Gamma}^\text{T})^{-1}\right]
\label{eq:N07b}.
\end{equation}

We need to give the specific form of %the diffusion matrix
${\bf D}$ for the calculation of the matrix $\mathbf J$.
For the  fluctuations in thermal equilibrium, the diffusion matrix is given by the fluctuation dissipation theorem ${\bf D}=k_\mathrm BT{\bf Q}_0$ \cite{Doi2013}, where $k_\mathrm B$ is the Boltzmann constant and $T$ is the system temperature.
 In contrast, the fluctuations in the active filament considered here %in this Letter 
 are the active fluctuations generated by the internal activity.
However, the universal properties that identify the active fluctuation have not been established. Here, we  employ the effective temperature $T^\mathrm {eff}$ \cite{Lau2009, Morozov2010, Loi2011, Levis2015} to determine the diffusion matrix of the active fluctuation, as a simple model.
%Although only the effective temperature cannot explain all the properties of active fluctuation, this concept can be considered as the most straightforward approach.
In fact, the effective temperature for a sperm cell is observed to be larger than the room temperature by an order of magnitude \cite{Ma2014}. 
With the effective temperature, the diffusion matrix becomes ${\bf D}=k_\mathrm BT^\mathrm{eff}{\bf Q}_0$, which can be obtained by replacing $T$ with $T^\mathrm{eff}$ in the fluctuation dissipation theorem.

 Because ${\bf J}$ is anti-symmetric,  when written as $J_{\alpha\beta}=J\epsilon_{\alpha\beta}$, we have
\begin{eqnarray}
J&=&-\frac{3L^6\gamma}{2\tau_n} \bigg[(\ell_1^3+\ell_3^3)\ell_2^3
+3\ell_1\ell_3(\ell_1^2+\ell_3^2)\ell_2^2\nonumber \\
&& +3\ell_1^2\ell_3^2(\ell_1+\ell_3)\ell_2+2\ell_1^3\ell_3^3\bigg]^{-1}
\label{eq:F12},
\end{eqnarray}
where we introduce viscosity drag coefficient $\eta_\|=2\pi\mu/\ln(2L/b)$, with $\mu$ and $b$ being the medium viscosity constant and the radius of the rod, respectively. We have also assumed that the drag ratio between the perpendicular and parallel components is anisotropic, $\eta_\perp=2\eta_\|$, and introduced a noise relaxation timescale, $\tau_n=\eta_\| L^3/ k_\mathrm BT^\mathrm{eff}$.
The final expression of the average swimming velocity only possesses an $\bm{e}_x$ component, denoted by $V_x$, and we obtain $V_x=-2FJ$, which is linearly proportional to $\gamma$ and the noise strength $ k_\mathrm BT^\mathrm{eff}$. In the case of three rods with equal lengths, $\ell_1=\ell_2=\ell_3=\ell$, the results are simply given by 
\begin{equation}
F=-\frac{5\ell}{81},~ J=-\frac{81\gamma}{16\tau_n}
\textrm{,~and}~
V_x=-\frac{5\gamma\ell}{8\tau_n}
\label{eq:F13}.
\end{equation}

%Though the mean rotational velocities are found to be zero, i.e., $\langle \theta\rangle=0$, due to the fluctuation, the angle value
The angle diffusion of the swimmer is obtained by calculating the squared angle displacement $\langle \theta^2\rangle$. We can estimate the angle $\theta$ as $$
\theta=\int_0^t A_{12\alpha}\dot{\alpha}_\alpha \,dt'=A_{12\alpha}\alpha_{\alpha}+\textrm{ higher-order terms}$$
for a small deformation, and it then follows that $\langle \theta^2\rangle\approx A_{12\alpha} C_{\alpha\beta}A_{12\beta}=o(t)$, indicating that the angle diffusion from the active fluctuation is negligible. We can also add thermal fluctuation in the system (\ref{eq:M02}), which would affect all the 5 degrees of freedom. Then, the odd-elastic swimmer can be represented as an active rotational Brownian particle with swimming velocity given by  (\ref{eq:F03}), and angle diffusion from the thermal noise. 

With a finite size of the fluctuation of $k_\mathrm {B}T^\mathrm{eff}$, a pusher filament can reach a stable limit cycle in a self-organized manner as demonstrated in Fig. \ref{fig:noise}(c), in which the swimmer is initially located at rest with a straight configuration, but once the shape exceeds the inner unstable limit cycle, it approaches the outer stable limit and exhibits self-organized periodic swimming (see Supplemental Movie). %The limit cycle found in the full simulation exhibits an ellipsoidal loop in the shape space, which has very similar to the ellipse of the probability current of the linear theory (Fig. \ref{fig:bifurcation}(b)). %Further, the strong correlation between the areal velocity and the swimming speed suggests that the formula (\ref{eq:F03}) is useful even for a non-small amplitude swimming. 

%\subsection{fluctuation around a limit cycle}
% We then try to write down the averaged swimming and the angle diffusion constant for stably swimming pusher swimmer under a small noise. We would need a new formulation similar to the Feynman path integral.

%\subsection{Example swimming with noise}

\section{General elastohydrodynamics}
By representing the force and torque balance equations via arbitrary degrees of freedom and their conjugate hydrodynamic force \cite{Doi2013}, we show that the
symmetric resistance matrix and the symbolic elastohydrodynamic equations, (\ref{eq:M02}) and (\ref{eq:M11b}), respectively, can be extended to a general linear elastic system. %material with a non-symmetric elastic matrix. 
 More precisely, the elastic matrix $\bf{K}$ can be an arbitrary positive-definite, $N\times N$ matrix, where $N$ is the number of degrees of freedom in shape space. Examples include $(N+1)$ spheres linked by $N$ arms and $(N+1)$ links connected by $N$ hinges, with the latter model being established as a coarse-grained representation of a continuous elastic filament \cite{Moreau2018, Walker2019d, walker2020efficient}. Hence, the results presented in this {\bl paper}, while mainly implemented for Purcell's swimmer here, are remarkably applicable to a wide class of low-Reynolds-number elastohydrodynamics.

Furthermore, we consider a general microswimmer at some steady state experiencing noise under the three assumptions of a) linearity of the shape dynamics ($\vec{\Gamma} = \bf{Q}_0 \bf{K}$), b) fluctuation dissipation theorem-type relationship with some effective temperature (${\bf D}=k_\mathrm BT^\mathrm{eff}{\bf Q}_0$), and c) null probability current ($\vec{j}=\bm{0}$), termed as the detailed balance relation \cite{Weiss2007}. We can then deduce that ${\bf J}={\bf 0}$ \textit{if and only if} the elasticity matrix $\bf{K}$ is symmetric. {\bl A formal proof is given in Appendix \ref{App:noise}}. 
As an important consequence, this formally demonstrates that an even-elastic swimmer can never exhibit directed locomotion from random noise, whereas every odd-elastic swimmer (non-symmetric $\bf{K}$) does.

%Overwhen ${\bf K}$ has non-zero anti-symmetric components, ${\bf J}$ cannot vanish because  it would violate the detailed balance relation \cite{Weiss2007}. Therefore, any general odd-elastic microswimmer around the equilibrium would exhibit noise-induced directed locomotion. 

%We can also formally demonstrate a general odd-elastic microswimmer around an equilibrium where the non-zero noise-induced swimming velocity originates from odd elasticity, because the condition ${\bf J}={\bf 0}$ violates the detailed balance relation \cite{Weiss2007}; conversely ${\bf J}$ cannot vanish if ${\bf \Gamma}$ has non-zero anti-symmetric components. In the odd elastic model with ${\bf \Gamma}={\bf Q}_0{\bf K}$, the detailed balance was broken directly by the odd elasticity.
%On the contrary, If detailed balance is hold, we can show that $\Omega= O$. First, $(\Omega D)^T=D\Omega^T=-D\Gamma^T+DC^{-1}D$. From the detailed balance relation, we find $(\Omega D)^T=\Omega D$. Remember that $\Omega C$ is anti-symmetric,  ???... thus $\Omega=O$??.
%Can we prove the opposite? i.e., $\Omega=O$ when the detailed balance is satisfied.

%In this latter case, i.e. w
%When the detailed balance is violated, the entropy $\sigma$ of the surrounding fluid is produced.
According to \cite{Weiss2007}, the entropy production rate is given by $T\langle \dot{\sigma} \rangle =-k_\mathrm BT^\mathrm{eff}\,\textrm{Tr}( \Gamma G)$ with the gain matrix ${\bf G}={\bf \Gamma CD}^{-1}-{\bf I}=2{\bf JD}^{-1}$.
Using ${\bf D}=k_\mathrm BT^\mathrm{eff}{\bf Q}_0$, we obtain $T\langle \dot{\sigma} \rangle =2\textrm{Tr}({\bf KJ})=-4\kappa_oJ>0$ and find that the odd part of the elasticity contributes to the entropy production and that the entropy production coincides with the average power obtained by the elasticity  $\langle \dot{W} \rangle=-K_{\alpha\beta}\langle \dot{\alpha}_{\alpha}\alpha_\beta\rangle$.
These results may be physically interpreted in the following way: 
nonconservative forces characterized by the odd elasticity generate work $W$ on the fluid; then, the fluid viscosity turns the work to heat, and the entropy of the fluid is produced.

%\section{Discussions and conclusions}
%{\it Conclusions.--}
\section{Discussion and conclusions}
%\subsection{Other boundary conditions}
%other boundary conditions.--
%Finally, we fix the $x$ position and the angle $\theta$ but let the swimmer move in the $y$ direction. The swimmer can move along the $y$ axis, and we use this boundary condition for a \emph{sliding swimmer}. This can be regarded as a model of molecular motors which move on a filament with changing its shape periodically driven by chemical reaction  (REF) .
%Due to the symmetry of the $x$ axis, even after the stable periodic waveform emerges for a pusher swimmer, it cannot generate net locomotion along the $y$ axis, but only oscillate in time. To generate non-zero sliding motion, it seems necessary to break this symmetry and this can be achieved by introducing an asymmetry with non-zero constant $K^o_0$ such as $K^o\epsilon_{\alpha\beta}\mapsto (K^o-K^o_0)\epsilon_{\alpha\beta}$.  Molecular motors such as  dynein and kinesin break the symmetry  (REF)
%\subsection{N-link swimmer}
% N-link swimmer.--
%In this study, with the simple Purcell swimmer, we formulate a general structure of elastohydrodynamic swimming of odd-elastic material in the low-Reynolds-number fluid. Through the hydrodynamic propulsion generated by the odd elasticity and the geometrical non-linearity, elastohydrodynamic buckling instability occurs for a fore-aft asymmetric swimmer and periodic self-organized swimming can be realized. This is extended to the case in the presence of noise in the shape space and general theory based on the gauge theory and probability current are proposed. 
 In this {\bl paper}, we describe our investigation of the elastohydrodynamics of a linear elastic material with a non-symmetric elastic matrix, with a focus on the analysis of Purcell's three-link swimmer. The odd elasticity,  represented by anti-symmetric parts of the elastic matrix, breaks the elastic reciprocity and leads to non-reciprocal deformation. With added internal fluctuation, we showed that the odd elasticity produces a net current in the shape space, hence generating net locomotion, and provided an explicit formula for the swimming velocity in (\ref{eq:F03}). For a pusher-type odd-elastic rod, time-periodic swimming is realized as a stable limit cycle, which is reachable from a straight configuration under a finite amount of noise, demonstrating self-organized swimming. 

This result suggests that some specific microswimming patterns emerging from internal activity in materials could be well captured by odd elasticity, as shown by the pusher-like beating of the odd-elastic three-link swimmer, although the limitations of this simple model are also clearly revealed by the absence of stable swimming for a puller-type rod. Further studies are needed to understand the microscopic origin of odd-elasticity %and the validity of the model for the internal activity in biological filaments such as flagella and cilia.
Moreover, this study also  advocates the potential of enforcing odd elasticity within artificial flexible microswimmers to generate autonomous motion, rather than explicitly prescribing the shape or using external controls.

The theoretical framework established in this study is applicable to a general swimmer with a large number of degrees of freedom.  % including three-dimensional motion of a general $N$-link swimmer which well approximates the elastic filament dynamics \cite{walker2020efficient}. %Nonetheless, the waveform of biological swimmers are often represented by a small number of dimensions, such by PCA modes, and described as a limit cycle with Gaussian noise in the PCA shape space \cite{Ma2014, Wan2014a, Ishimoto2017b}. Further studies including biological data could shed light on the non-reciprocal property of active elastohydrodynamics systems.
The function of many biological molecules depends on their shape, as seen in molecular machines and enzymes whose shape changes with fluctuations \cite{Ariga2018, Brown2019, Mugnai2020, Yasuda2021b, Ghosh2021}.
The swimming formula of noisy elastic material may be applied to these micromachines, but this is also left as future work. 

%In conclusion,} the current study provides general features of the microscopic noisy elastohydrodynamics and therefore will be useful in modeling, understanding, and designing biological and artificial active materials.

%\section*{Acknowledgements}
\section*{Acknowledgments}
K.I. acknowledges the Japan Society for the Promotion of Science (JSPS), KAKENHI for Young Researchers (Grant No. 18K13456) and Transformative Research Areas A (Grant No. 21H05309) and the Japan Science and Technology Agency (JST), PRESTO Grant (No. JPMJPR1921). C.M. is a JSPS International Research Fellow (PE20021). K.Y. acknowledges support by a Grant-in-Aid for JSPS Fellows (Grant No. 21J00096) from the JSPS. K.I., C.M., and K.Y. were partially supported by the Research Institute for Mathematical Sciences, an International Joint Usage/ Research Center located at Kyoto University. The authors thank anonymous referees for their useful feedback that helped us improve the manuscript.

\appendix

\begin{appendix}
% We'll move the Appendices to Supplementary before submission

\section{Symmetric properties of the resistance matrix}
\label{App:MatM}

We denote the positions of ends of the rods in the body-fixed coordinates as $\bm{r}_0, \bm{r}_1, \bm{r}_2, \bm{r}_3$. From the definition of the body-fixed coordinates, $\bm{r}_0=\bm{0}$, $\bm{r}_1=\ell_1\bm{e}_{x0}$, $\bm{r}_2=\bm{r}_1+\ell_2(\cos\alpha_1\bm{e}_{x0}+\sin\alpha_1\bm{e}_{y0})$, and $\bm{r}_3=\bm{r}_2+\ell_2(\cos(\alpha_1+\alpha_2)\bm{e}_{x0}+\sin(\alpha_1+\alpha_2)\bm{e}_{y0})$.
From the linearity, the two-dimensional surface velocity of a Purcell swimmer in the body-fixed coordinates at the position $\bm{r}$ can then be represented in matrix form, using the state vector $\bm{z}$, as
\begin{equation}
    \bm{v}={\bf H}\dot{\bm{z}}
    \label{eq:S01}.
\end{equation}
The entries of the matrix ${\bf H}={\bf H}(\bm{r};\bm{z})$ are given by
\begin{equation}
    {\bf H}=\begin{pmatrix}
     1 & 0 & -r_0\sin\theta_0 & -g_1(\bm{r})r_1\sin\theta_1 &-g_2(\bm{r})r_2\sin\theta_2\\
     0 & 1 & r_0\cos\theta_0 & g_1(\bm{r})r_1\cos\theta_1 & g_2(\bm{r})r_2\cos\theta_2
    \end{pmatrix}
    \label{eq:S02},
\end{equation}
where we have introduced the lengths of vectors $r_0=|\bm{r}-\bm{r}_0|$, $r_1=|\bm{r}-\bm{r}_1|$, and $r_2=|\bm{r}-\bm{r}_2|$; angles from the $\bm{e}_{x0}$ axis as
$\theta_0=\textrm{arg}(\bm{r}-\bm{r}_0)$, 
$\theta_1=\textrm{arg}(\bm{r}-\bm{r}_1)$, and $\theta_2=\textrm{arg}(\bm{r}-\bm{r}_2)$;
and functions $g_1(\bm{r})$ and $g_2(\bm{r})$ as
\begin{equation}
g_1(\bm{r})=\begin{cases} 0 & (\bm{r} \textrm{~~is on the first rod}) \\
1 & (\bm{r} \textrm{~~is on the second and third rods})
\end{cases} 
    \label{eq:S02a}
\end{equation}
and
\begin{equation}
g_2(\bm{r})=\begin{cases} 0 & (\bm{r} \textrm{~~is on the first and second rods}) \\
1 & (\bm{r} \textrm{~~is on the third rod})
\end{cases} 
    \label{eq:S02b}.
\end{equation}

The surface traction force $\bm{f}$ defines the conjugate  hydrodynamic force vector $\bm{h}$  by the integral over the swimmer surface \cite{Doi2013}
\begin{equation}
\bm{h}= \int_S \bm{f}^\text{T}{\bf H}\,\textrm{d}S
    \label{eq:S03}.
\end{equation}
By direct calculation, we obtain 
$\bm{h}=(F_{x0}, F_{y0}, T_z, T_{1z}, T_{2z})^\textrm{T}$. Here, $F_{x0}$ and  $F_{y0}$ are the total hydrodynamic force along the $\bm{e}_{x0}$ and $\bm{e}_{y0}$ axes, respectively, and $T_z$ is the total hydrodynamic torque. $T_{1z}$ and $T_{2z}$ are internal torque around the points $\bm{r}_1$ and $\bm{r}_2$, respectively, and these should be balanced by the elastic torque. If we write down the force and torque balance equations for the vector $\bm{h}$, we obtain the elastohydrodynamic equation
\begin{equation}
    -{\bf M}\dot{\bm{z}}={\bf L}\bm{z}
    \label{eq:S04},
\end{equation}
as in the main text, where ${\bf M}$ is the resistance matrix. If the resistance matrix is introduced between the generalized velocity and its conjugate force, the resistance matrix found to be symmetric and positive-definite by the  Lorentz reciprocal relation \cite{Doi2013}.

As in the main text, we introduce lengthscale $L=\ell_1+\ell_2+\ell_3$ and viscosity drag coefficient $\eta_\|=2\pi\mu/\ln(2L/b)$ with $\mu$ and $b$ being the medium viscosity constant and radius of the rod, respectively.  We also assume the drag anisotropy ratio between the perpendicular and parallel components, $\eta_\perp=2\eta_\|$. %, and noise relaxation timescale, $\tau_n=\eta_\| L^3/ k_\mathrm BT^\mathrm{eff}$. 
The expression for $Q_0$ is then given by
\begin{eqnarray}
    Q_{0,11}&=&\frac{6}{\eta_\| L^3}\times
    \frac{(\ell_1+\ell_2)^3(\ell_1+\ell_2+\ell_3)^4}{\ell_1^3\ell_2^2(4\ell_2(\ell_1+\ell_2+\ell_3)+3\ell_1\ell_3)}
\\  Q_{0,22}&=&  \frac{6}{\eta_\| L^3}\times\frac{(\ell_2+\ell_3)^3(\ell_1+\ell_2+\ell_3)^4}{\ell_2^2\ell_3^3(4\ell_2(\ell_1+\ell_2+\ell_3)+3\ell_1\ell_3)}
\\  Q_{0,12}&=&Q_{0,21}=-\frac{3}{\eta_\| L^3}\times 
 L^3\bigg[3\ell_2^3+6(\ell_1+\ell_3)\ell_2^2 \nonumber \\
&&+(3\ell_1^2+8\ell_1\ell_3+3\ell_3^2)\ell_2+2\ell_1\ell_3(\ell_1+\ell_3) \bigg]\nonumber \\
&& /\big[ \ell_1\ell_2^2\ell_3(4\ell_2(\ell_1+\ell_2+\ell_3)+3\ell_1\ell_3) \big].
\end{eqnarray}

\section{Influence of the swimmer's geometry}
\label{App:geom}

\begin{figure}[!t]
\begin{center}
\includegraphics[width=8.0cm]{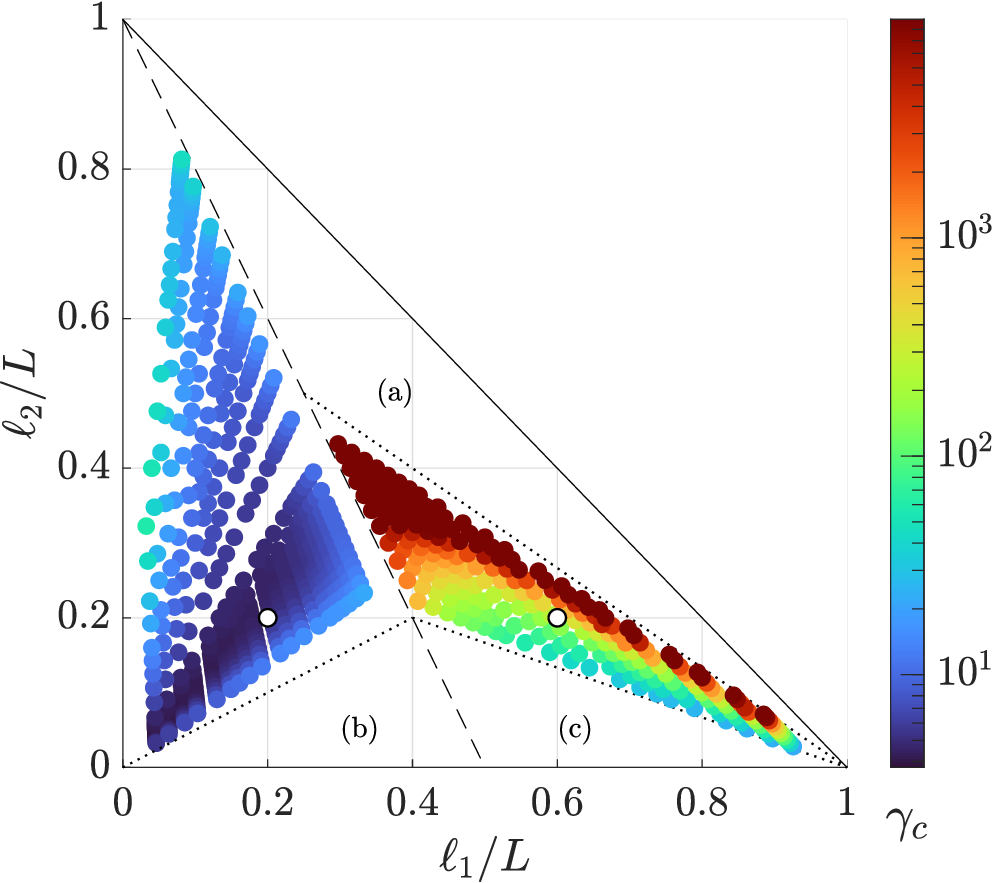}
\caption{Bifurcation value $\gamma_c$ with respect to the ratios $\ell_1/L$ and $\ell_2/L$. The dashed line separates the $\ell_1<\ell_3$ region on the left (puller case) from the $\ell_1>\ell_3$ region on the right (pusher case). In areas (a), (b), and (c), delineated by indicative dotted lines, no bifurcation can be observed. White dots represent the values used in Fig. \ref{fig:bifurcation}.}
\label{fig:lengths}
\end{center}
\vspace*{-0.75cm}
\end{figure}

In this section, we discuss the influence of the segment lengths $\ell_1, \ell_2, \ell_3$ on the swimming behavior, particularly on the value $\gamma_c$ at which the bifurcation displayed in Fig. \ref{fig:bifurcation} occurs. 
As above, we define the lengthscale $L=\ell_1+\ell_2+\ell_3$ and study the variation of $\gamma_c$ with respect to the non-dimensional ratios $\ell_1/L$ and $\ell_2/L$, as shown in Fig. \ref{fig:lengths}. For positive values of $\gamma$, the swimmer behaves as a puller (resp. pusher) if $\ell_1<\ell_3$ (resp. $\ell_1>\ell_3$), which coincides with the region on the left (resp. right) of the dashed line on Fig. \ref{fig:lengths}. The colored dots cover the areas where a bifurcation similar to the ones presented in Fig. \ref{fig:bifurcation} can be observed, showing that this bifurcation phenomenon holds for a wide array of swimmer geometries. The white dots indicate the values used in Fig. \ref{fig:bifurcation}. Strikingly, the bifurcation occurs for much larger values of $\gamma$ in the pusher case, as indicated by the colors in Fig. \ref{fig:lengths}: $\gamma_c$ remains less than $10^2$ in the  puller case, while it ranges between $10^2$ and more than $10^4$ in the pusher case. In regions marked (a) and (b), all the orbits converge to the stable equilibrium 0 -- however, the convergence speed is extremely slow. In region (c), the nonzero stable cycle observed for the pusher becomes unstable, with the orbits then converging to a nonphysical stable cycle (with values of $\alpha_i$ greater than $\pi$). 

\section{Noisy swimmer around the equilibrium}
\label{App:noise}

Here, we consider an $n$-dimensional general swimmer with an $N$-dimensional shape space; in particular, for the Purcell swimmer, $n=2$ and $N=2$. We write the generator for the $n$-dimensional Euclidean group $\mathcal{A}$, which can be represented as an $(n+1)\times(n+1)$ matrix. The swimming velocity with a small deformation can be expanded as
\begin{equation}
\mathcal{A}(\bm{\alpha})=\mathcal{A}_\alpha(\bm{0})\dot{\alpha}_\alpha+\frac{1}{2}\left(\mathcal{G}_{\alpha\beta}(\bm{0})+\mathcal{F}_{\alpha\beta}(\bm{0})\right)\,\alpha_\alpha\dot{\alpha}_\beta
\label{eq:N01},
\end{equation}
where $\mathcal{G}_{\alpha\beta}$ and $\mathcal{F}_{\alpha\beta}$ indicate the symmetric and anti-symmetric part of the second-order term, respectively, and the latter is identical to the strength of the gauge field or the curvature of the gauge field. 

We consider the Langevin equation for the dynamics around the origin of the shape space, given by
\begin{equation}
\dot{\bm{\alpha}}=-{\bf \Gamma}\bm{\alpha}+\bm{\xi}(t)
    \label{eq:N02},
\end{equation}
with the zero-mean Gaussian noise $\xi_\alpha(t)$. Here, the $N\times N$ matrix ${\bf \Gamma}$ is constant in time and positive-definite, representing deterministic dynamics, which is linearly stable around the origin of the shape space.  For the Purcell swimmer in the main text, ${\bf \Gamma}={\bf Q}_0 {\bf K}$.
The Gaussian noise satisfies
\begin{equation}
\langle \xi_\alpha(t)\xi_\beta(t') \rangle=2D_{\alpha\beta}\delta(t-t')
    \label{eq:N03},
\end{equation}
where the brackets indicate ensemble average, $\delta(t)$ is the Dirac delta function, and the diffusion matrix ${\bf D}$ is symmetric and positive-definite. %We assume the diffusion matrix is related to the mobility sub-matrix in the same from as the fluctuation-dissipation theorem with an effective temperature of ${\bf D}=\kappa_n{\bf Q}_0$, where $\kappa_n$ is the amount of noise with a physical unit of energy and assumed to be sufficiently larger than the room temperature $\kappa_n\gg k_BT$. 
 Hereafter, we do not explicitly indicate that the evaluation point is at $\bm{\alpha}=\bm{0}$.

By plugging  (\ref{eq:N03}) into  (\ref{eq:N02}) and noting that the equal-time correlation between the shape and the noise can be obtained by  $\langle \alpha_{\alpha}\xi_\beta\rangle=D_{\alpha\beta}$, the average swimming velocity becomes 
\begin{eqnarray}
\langle \mathcal{A} \rangle
&=&\frac{1}{2}\left( \mathcal{G}_{\alpha\beta} + \mathcal{F}_{\alpha\beta} \right)\langle \alpha_\alpha\dot{\alpha}_\beta\rangle \nonumber \\
&=&-\frac{1}{2}\left( \mathcal{G}_{\alpha\beta} + \mathcal{F}_{\alpha\beta}\right)\left( \Gamma_{\beta\gamma}C_{\gamma\alpha} -D_{\alpha\beta}\right)
\label{eq:N04},
\end{eqnarray}
because the first-order term vanishes, that is, $\langle \dot{\alpha}_\alpha \rangle=0$. Here we introduced the shape covariance matrix,  $C_{\alpha\beta}=\langle  \alpha_\alpha \alpha_\beta\rangle$, which obeys the
%We may rewrite
%\begin{equation}
%  \langle \mathcal{A} \rangle =-\frac{1}{2}\textrm{Tr}(\mathcal{F}^\ast\Gamma C)  
%\label{eq:N05},
%\end{equation}
%where the trace is taken over the shape indices. The covariance matrix in the Langevin equation (\ref{eq:N02}) obeys to the
Lyapunov equation \cite{Weiss2007},
\begin{equation}
{\bf \Gamma C} + {\bf C\Gamma}^\textrm{T}=2{\bf D}    
\label{eq:N06-app},
\end{equation}
with superscript $\textrm{T}$ denoting transpose of the matrix, and the formal solution may be written as
\begin{equation}
{\bf C}=2\int_{-\infty}^0 e^{{\bf \Gamma} t}\,{\bf D} \,e^{{\bf \Gamma}^\textrm{T} t}dt    
\label{eq:N07}.
\end{equation}
% Fokker-Planck equation and probability flux\\
The time-dependent probability distribution function in the shape space $p(\bm{\alpha},t)$ can be obtained from the Fokker-Planck equation associated with (\ref{eq:N02}), i.e., $\partial p/\partial t +\partial j_\alpha/\partial \alpha_\alpha=0$ with the probability flux $j_\alpha$ given by
\begin{equation}
j_\alpha=-p\,\Gamma_{\alpha\beta}\,\alpha_\beta -D_{\alpha\beta}\frac{\partial p}{\partial \alpha_\beta}
\label{eq:N08}.
\end{equation}
The steady-state probability distribution function is the Gaussian function,
\begin{equation}
p(\bm{\alpha})=\frac{1}{(2\pi)^{N/2}\sqrt{\det{{\bf C}}}}\exp{\left[-\frac{1}{2}\bm{\alpha}^{\textrm{T}}{\bf C}^{-1}\bm{\alpha}\right]}
\label{eq:N09},    
\end{equation}
and plugging this expression into  (\ref{eq:N08}) leads to the steady-state probability current, $\bm{j}={\bf \Omega}\bm{\alpha}p$, where the matrix ${\bf \Omega}$ is defined as
\begin{equation}
{\bf \Omega}=-{\bf \Gamma}+{\bf DC}^{-1}    
\label{eq:N10},
\end{equation}
where ${\bf \Omega}$ may be interpreted as the matrix of rotational velocity of the probability current in the shape space and the vector ${\bf \Omega}\bm{\alpha}$ represents the shape space velocity. At steady state, by $\partial j_\alpha /\partial \alpha_\alpha=0$, substituting  (\ref{eq:N09}) yields the traceless property of the matrix, $\textrm{Tr}\,{\bf \Omega}=0$.
%\begin{figure}[!t]
%\begin{center}\includegraphics[width=6cm]{fig_L1_g1.eps}
%\caption{An illustration of the steady probability distribution shown by color contour and the probabilistic current vector in the shape space depicted by blue arrows. The solution for the equal-sized rod swimmer with $\gamma=1$ is shown.}
%\label{fig:current}
%\end{center}
%\vspace*{-0.75cm}
%\end{figure}

Let us write 
\begin{equation}
{\bf J}=\frac{1}{2}{\bf \Omega C}=\frac{1}{2}({\bf D}-{\bf \Gamma C})
\label{eq:N10a},
\end{equation}
which may be interpreted as the areal velocity of the probability current in the shape space. By the definition of ${\bf J}$, its transpose becomes $2{\bf J}^{\textrm{T}}={\bf C\Omega}^{\textrm{T}}={\bf C}({\bf C}^{-1}{\bf D}-{\bf \Gamma}^{\textrm{T}})$, and 
from the Lyapunov equation (\ref{eq:N06}), it follows that ${\bf J}^{\textrm{T}}=-{\bf J}$. Thus, we conclude that ${\bf J}$ is anti-symmetric.
With areal velocity matrix ${\bf J}$, the average swimming velocity (\ref{eq:N04}) can be finally derived:
\begin{equation}
\langle \mathcal{A}\rangle =\textrm{Tr}\left((\mathcal{G}+\mathcal{F}){\bf J}\right)
=\textrm{Tr}(\mathcal{F}{\bf J})  
\label{eq:N11},
\end{equation}
where the trace is taken over the shape indices.
In the second equality, we used $\textrm{Tr}(\mathcal{G}{\bf J})=0$, because $\mathcal{G}$ is symmetric and ${\bf J}$ is anti-symmetric in the indices of the shape space.

For simplicity, we first consider the $N=2$ case as in the main text. We rewrite $(\mathcal{F}_{\alpha\beta})_{ij}=F_{ij\alpha\beta}=(\sqrt{\det{\mathcal{F}}})_{ij}\epsilon_{\alpha\beta}$ with an $(n+1)\times(n+1)$ matrix $\sqrt{\det{\mathcal{F}}}$, noting that the determinant is taken for the shape space labels.
By direct calculation, we have $\textrm{Tr}(\mathcal{F}\,{\bf J}) =-2\sqrt{\det{\mathcal{F}}}\sqrt{\det{{\bf J}}}=-2\sqrt{\det{\mathcal{F}}}\sqrt{\det{{\bf C}}}\sqrt{\det{{\bf \Omega}}}$.  From the zero trace of ${\bf \Omega}$, the eigenvalues of the matrix are simply $\nu=\pm i\,\sqrt{\det{{\bf \Omega}}}$, which are pure imaginary. The average swimming velocity is therefore written as 
\begin{equation}|\langle \mathcal{A}\rangle|=2\sqrt{\det{\mathcal{F}}}\sqrt{\det{{\bf C}}}\, |\nu|
\label{eq:N12},
\end{equation}
which successfully generalizes the finding in the three-sphere model [Eq. (19) of Ref. \cite{Yasuda2021}]. The form in (\ref{eq:N12}) shows that the average velocity is represented by the products of a shape-dependent geometrical factor, the explored area in the shape space, and the speed of the rotational probability flux. The product of the latter two indicates the areal velocity.

Formula (\ref{eq:N12}) is easily extended into a general shape space with $N$ dimensions. The anti-symmetric matrix, ${\bf J}$, can be block-diagonalized by an orthogonal matrix,
\begin{equation}
\begin{pmatrix}
 \tilde{{\bf J}}^{(1)} &  & O & O\\
  & \ddots &  & \vdots \\
 O & & \tilde{{\bf J}}^{(d)}  & \vdots\\
 O & \cdots  & \cdots & O
 \end{pmatrix}
 \label{eq:N13},
\end{equation}
in which the $2\times 2$ matrices $\tilde{{\bf J}}^{(1)}\cdots,\tilde{{\bf J}}^{(d)}$ are all real anti-symmetric with eigenvalues all pure imaginary, where $d=[N/2]$ is the integer part of $N/2$. With positive real numbers $\omega_1, \cdots,\omega_d$, the matrices are represented as $\tilde{J}^{(1)}_{\alpha\beta}=i\omega_1\epsilon_{\alpha\beta}$. Similarly, with the same orthogonal transformation,
\begin{equation}
\begin{pmatrix}
 \tilde{\mathcal{F}}^{(1)} &  & \ast & \ast\\
  & \ddots &  & \vdots \\
 \ast & & \tilde{\mathcal{F}}^{(d)}  & \vdots\\
 \ast & \cdots  & \cdots & \ast
 \end{pmatrix}
 \label{eq:N14},
\end{equation}
where asterisks denote arbitrary entries. Here, $\tilde{\mathcal{F}}^{(1)}\cdots,\tilde{\mathcal{F}}^{(d)}$ are real anti-symmetric with respect to the shape indices, and we again rewrite these as $(\tilde{\mathcal{F}}^{(1)})_{\alpha\beta}=\tilde{F}^{(1)}_{ij\alpha\beta}=(\sqrt{\det\tilde{\mathcal{F}}^{(1)}})_{ij}\epsilon_{\alpha\beta}$. The average swimming velocity is then simplified in the form
\begin{equation}
|\langle \mathcal{A}\rangle|=\sum_{q=1}^{d}|\textrm{Tr}(\tilde{\mathcal{F}}^{(q)}\tilde{{\bf J}}^{(q)})|=\sum_{q=1}^d |\omega_q|\sqrt{\det\tilde{\mathcal{F}}^{(q)}}
\label{eq:N15}.
\end{equation}
The average swimming velocity is decomposed into the contributions from the shape subspace and represented as the product of field strength and areal velocity in each 2D sub-space. In the case with 3D shape space ($N=3$), the probability current lies in the 2D plane in 3D space. When $N=4$, the dynamics in the shape space are then decomposed into two separated 2D planes.

Finally, we demonstrate that the non-zero noise-induced swimming velocity originates from the non-symmetric property of the matrix ${\bf K}$, under the assumptions of  a) linear  shape dynamics ($\vec{\Gamma} = \bf{Q}_0 \bf{K}$), b) fluctuation dissipation theorem-type relationship with some effective temperature (${\bf D}=k_\mathrm BT^\mathrm{eff}{\bf Q}_0$) and c) null probability current ($\vec{j}=\bm{0}$).

The null probability current leads to ${\bf \Omega}={\bf 0}$ from Eq. (\ref{eq:N10}) and we therefore have ${\bf \Gamma}={\bf D}{\bf C}^{-1}$. Substituting this into the Lyapunov equation (\ref{eq:N06}) and eliminating ${\bf C}$, we obtain ${\bf \Gamma} {\bf D}-{\bf D}{\bf \Gamma}^{\textrm{T}}={\bf 0}$.  
With assumptions (a) and (b), we then obtain ${\bf D}({\bf K}-{\bf K}^\textrm{T}){\bf D}={\bf 0}$. Thus the null probability current is equivalent to the symmetric property of the elastic matrix ${\bf K}$. Conversely, this indicates that a non-symmetric ${\bf K}$ generates non-zero ${\bf \Omega}$ and non-zero ${\bf J}$, yielding the non-zero swimming velocity from formula (\ref{eq:N11}). 

%\begin{figure}[b]
%\begin{center}
%\includegraphics[width=7.0cm]{figS2.eps}
%\caption{Snapshot of Supplemental Movie.}\label{fig:sm}
%\end{center}
%\vspace*{-0.75cm}
%\end{figure}

%
%\section{Caption of Supplemental Movie}
%\label{App:caption}
%A simulation movie of Purcell's swimmer undergoing shape fluctuation with the positions of the end of the left-most rod is traced over time. A pusher swimmer is considered with the model parameters $(\ell_1/L, \ell_2/L, \ell_3/L)=(0.2, 0.2, 0.6)$ and $\gamma=-300$, where $L=\ell_1+\ell_2+\ell_3$ is the total length of the rods. This set of parameters exhibits the same dynamics as in Fig. 1(b,c), and Fig. \ref{fig:bifurcation} but with opposite swimming direction. For the amount of noise, we used $k_\mathrm{B}T^\mathrm{eff}/\kappa_e=500$. The odd-elastic swimmer was initially at rest with a straight configuration. The dynamics are separated into three phases. (1) The swimmer moves slightly towards the right with a small amplitude (blue), until the shape exceeds the inner unstable limit cycle. (2) The beating wave is then amplified (orange) and reaches stable swimming (yellow).
\end{appendix}

\bibliography{library}

\end{document}